\documentstyle[12pt]{article}
\newcommand{\alt}{\mathbin{\lower 3pt\hbox
   {$\rlap{\raise 5pt\hbox{$\char'074$}}\mathchar"7218$}}}
\newcommand{\agt}{\mathbin{\lower 3pt\hbox
   {$\rlap{\raise 5pt\hbox{$\char'076$}}\mathchar"7218$}}}

\begin{document}
\setcounter{footnote}{0}
\setcounter{equation}{0}
\setcounter{figure}{0}
\setcounter{table}{0}
\vspace*{5mm}

\begin{center}
{\large\bf Anderson transition: numerical vs analytical results \\
(comment on the review article by P.Markos cond-mat/0609580) }

\vspace{4mm}
I. M. Suslov \\
P.L.Kapitza Institute for Physical Problems,
\\ 119337 Moscow, Russia \\
\vspace{6mm}

\begin{minipage}{135mm}
{\rm {\quad} In the recent review article, P.Markos admits that
practically all numerical results on the critical behavior
near the Anderson transition are in conflict with analytical
expectations, but no serious discussion of this fact is given.
The aim of the present comment is to give an analysis of the
arising situation.
}
\end{minipage} \end{center}

\vspace{6mm}

The recent paper by P.Markos \cite{1} provides the extensive review on
numerical investigations of  Anderson localization. Such
detailed review is quite valuable and long expected. Its value
is slightly diminished by the fact that the author is too
anxious to present his own results, even in the cases when the results
of other authors are more advanced. No preference is given to
the papers where record system sizes are achieved \cite{2,3,4,5};
no attention is given to results, which contradict to a conventional
paradigm \cite{6,7}; not all characteristics of the wave functions
studied in the literature  \cite{8} are discussed.

It is admitted in Sec.\,13.3 of \cite{1} that practically all numerical
results on the critical behavior near the Anderson transition are
in contradiction with the analytical predictions, but no
serious discussion of this circumstance is given. In fact, a situation
with numerical algorithms is rather serious, and the aim
of the present comment is to analyse this situation.

{\qquad\qquad\qquad\qquad\qquad--------------------------}

Fig.\,62 in \cite{1} presents the dependence of the critical
exponent $\nu$ of the correlation length on the space
dimensionality $d$ and its comparison with predictions of the
self-consistent theory by Vollhardt and W$\ddot o$lfle \cite{9}
(see also papers \cite{9a}). Such comparison is rather instructive
since the results of \cite{9} summarize in the compact form all
theoretical expectations. Nevertheless, the original version of
the theory \cite{9} is rather crude\,\footnote{\,The evident
drawbacks of the self-consistent theory \cite{9} are the crude
method of solving the Bethe--Salpeter equation, violation of the
Ward identity and neglection of the possible spatial dispersion of
the diffusion coefficient. These drawbacks were removed in the
symmetry approach of the paper \cite{10}: if only evident symmetry
of the system is taken into account and situation of the general
position (compatible with this symmetry) is considered, then the
results of \cite{9} are reproduced. The arguments on the absense
of the hidden symmetry can be also given \cite{10}, but these arguments
cannot be considered as indisputable. }
and contradiction with it does
not provide a serious argument against the numerical methods.
However, there are two fundamental contradictions that should be
discussed.
\vspace{3mm}

(A) Numerical results are in conflict with the analytical prediction
$\nu=1/\epsilon$ at $\epsilon\to 0$, obtained for a space dimensionality
$d=2+\epsilon$ \cite{10a}. The reference on the specific result by Hikami
given in \cite{1} (see Eqs.\,193,\,194) is disputable: the theory for
$d=2+\epsilon$ surely needs some modification due to instability
of the renormalization group caused by the high-gradient catastrophe
\cite{11}. However, the result $\nu=1/\epsilon$ for $\epsilon\to 0$
is valid for an arbitrary $\beta$-function of the form
$$
\beta(g)=\epsilon +\frac{A_1}{g} +\frac{A_2}{g^2}+\frac{A_3}{g^3}
+\ldots
\eqno(1)
$$
with $A_1<0$ ($g$ is the Thouless conductance), i.e. in any
variant of the theory compatible with  general philosophy of
one-parameter scaling \cite{12}\,\footnote{\,The negative sign of
$A_1$ is a basis of the weak localization theory \cite{12a} and
has numerous experimental confirmations.}
Contradiction with the result $\nu=1/\epsilon$ is possible, only if
the one-parameter scaling hypothesis is rejected; but then
the treatment of the raw numerical data becomes self-contradictory,
being entirely based on this hypothesis.

\vspace{3mm}

(B) The dependence of $\nu$ on $d$ (Fig.\,62 in \cite{1}) looks
perfectly smooth in the interval $2<d\le 5 $: such behavior is
in direct contradiction with the Bogolyubov theorem on
renormalizability of the $\varphi^4$ field theory \cite{36},
which is mathematically
equivalent to the problem of an electron in the Gaussian random
field \cite{13,14,15,16}. The $\varphi^4$ theory is renormalizable
for $d\le 4$ and nonrenormalizable for $d>4$, so $d=4$ is a singular
point on the $d$-axis. It is expected (though it is not a theorem)
that $d=4$ is an upper critical dimension\,\footnote{\,A situation
with the upper critical dimension is perfectly clear for the problem
of density of states. Simplification of theory for $d>4$ was demonstrated
and $(4-\epsilon)$-dimensional theory was developed in the series
of papers \cite{16a} (see also review \cite{16}). },
where $d$-dependence of the
critical exponents has a cusp, while for $d>4$  they are independent
on $d$ and
equil to their mean field values ($\nu=1/2$ in the present case). Nothing of
the kind is observed numerically. Furthermore, nonrenormalizability for $d>4$
makes existence of scale invariance {\it absolutely impossible} (due to
impossibility to exclude microscopic length scales from any quantity). In spite
of this, the numerical algorithms seems to have no difficulties in
interpretation of results in terms of one-parameter scaling, and it
demonstrates their quality. In fact, one can derive from the Vollhardt and
W$\ddot o$lfle theory \cite{9} or from two-parameter scaling \cite{17} that the
Thouless conductance is not stationary in the critical point for dimensions
$d\ge 4$; so the conventional condition for a critical point (see Fig.\,61 in
\cite{1}) becomes invalid (see discussion in \cite{18}[Sec.\,4.2]).

\vspace{5mm}

New problems in numerical algorithms were discovered recently
in studies of the second moments for a solution of the Cauchy
problem for the Schr$\ddot o$dinger equation in quasi-1D systems
\cite{18,19,20}. Roughly  half of numerical papers use finite-size
scaling for the minimal Lyapunov exponent $\gamma_{min}$
\cite{1}[Sec.\,12.1], related with the growth of the typical
value of the Cauchy solution.
Analytical calculation of $\gamma_{min}$ is possible only under
rather restrictive assumptions \cite{21,22}. On the other hand,
a complete analytical investigation is possible for the minimal
exponent $\beta_{min}$, characterizing a growth of the second
moments \cite{20}. Inequality $\beta_{min}\ge 2 \gamma_{min}$
can be rigorously established, while the order of magnitude relation
$\beta_{min}\sim \gamma_{min}$ is expected in the typical physical
situation; the latter relation is valid for weak \cite{21} and strong
\cite{22} disorder and confirmed by extensive numerical studies
\cite{23}. From viewpoint of  general scaling philosophy the
use of $\beta_{min}$ or  $\gamma_{min}$ is practically equivalent.
The following becomes clear as a result of such studies.
\vspace{3mm}

(C) In $2D$ systems, a conventional finite-size scaling approach
based on $\beta_{min}$ leads to unumbiguous conclusion on  existence
of  the Kosterlitz-Thouless type transition between exponential
and  power law localization, and absense of one parameter scaling
for $\beta_{min}$ \cite{20}. Inequality
$\beta_{min}\ge 2 \gamma_{min}$ makes it possible to establish
validity of both conclusions also for  finite-size scaling based
on $\gamma_{min}$. There exists the possibility to restore
one-parameter scaling, but it requires essential modification
of the conventional algorithm: one should use some effective exponent
$\gamma_{eff}$ instead $\gamma_{min}$. After such modification,
the $2D$ phase transition is predicted not for all systems.

These considerations are confirmed by Fig.\,37 in \cite{1},
where absense of scaling for $\langle g\rangle$ is clearly
visible: different dependences cannot be reduced to a single
curve by a scale transformation. The analogous behavior
can be surely found for other quantities, but the data for weak
disorder ($W<4$)  are practically never presented in numerical
papers.
\vspace{3mm}

(D) For $d>2$, a finite-size scaling based on $\beta_{min}$ leads
to the values of critical disorder, essentially different from those
obtained in numerical experiments \cite{18}.\,\footnote{\,The
results for critical disorder, obtained in \cite{18,20} and \cite{19}
by the essentially different methods, coincide for $d=2$ and $d\ge 4$
(not for $d=3$), but their interpretation is different. }
Two different interpretations of this fact are possible:

(a) Contemporary numerical experiments give incorrect results for
the critical point. It is related with the principal inapplicability
of one-parameter scaling for $d\ge 4$, while for $d=3$ a large
length scale $L_0$ exists, so that a true critical behavior
can be found only for systems with $L\agt L_0$ (see \cite{18}[Sec.\,4.2]
for details).

(b) If we accept validity of the conventional numerical results,
then we should assume that the point of the Anderson transition is
expanded into the band of the critical states \cite{18}. Surprisingly,
such possibility has  direct numerical confirmations
\cite{6,7}.\,\footnote{\,If they are not artifacts related with
existence of the large scale $L_0$.}
 Realization of such possibility signifies that all
existing analytical approaches are incorrect. In particulary, it
means invalidity of the one-parameter scaling theory \cite{12}, and
interpretation of practically all numerical experiments becomes
internally inconsistent.

One can see, that (a) and (b) have approximately the same consequences
for numerical algorithms. The given alternative is probably resolved in
favour of (a) in the recent paper \cite{23a}. Indeed,
all dependences in Fig.\,2 of \cite{23a} are practically linear for
$L<30$ in accordance with the old results by Schreiber \cite{7};
so the crossover between Fig.\,5,a and Fig.\,5,b of \cite{18}
occurs at $L\sim 30$, and a large length scale $L_0\sim 30$
appears in the 3D Anderson model.

\vspace{5mm}

One more problem is related with the critical distribution
of conductance.
\vspace{2mm}

(E) Figs.\,39,\,40,\,43,\,44,\,46,\,52 in \cite{1} show the
distribution of conductance  $p_c(g)$ in the critical point:
it has a singularity at $g=1$ and the rapid decrease for
large $g$. Such behavior is quite
understandable (see \cite{1}[Sec.\,10]) for quasi-1D systems
(with the size $L^{d-1}\times L_z$ and sufficiently large $L_z$),
but its validity for the real $d$-dimensional systems
($L\sim L_z$) seems doubtful. The singularity at $g=1$ looks
incredible\,\footnote{\,There are no phase transitions in finite
systems according to the Lee and Yang theorem. Analogously,
there is no ground for any other singularities in them. If
$p_c(g)$ is independent of $L$, then it can be calculated for
finite $L$.}, and the decrease more quicker than exponent
for large $g$ contradicts both to the direct
renormalization group analysis for high moments of $g$ \cite{24} and to
the attempts of reconstruction of the whole distribution $p_c(g)$
\cite{25}.  The second defect may be attributed to not sufficiently large
system size $L$, but it is claimed in \cite{1} that the
distribution is stationary and independent on $L$
(as a result, validity of one-parameter scaling is declared for
the whole distribution). Such thing is impossible, if we have any
belief in analytical theory. It looks more likely, that  formulas
for multi-channel localization are used beyond the range of their
applicability. It is not evident in the general case
that the Landauer conductance measured in numerical experiments
coincides with the Kubo conductance treated in analytical papers.
Indeed, the method used for calculation of conductance fails outside the
unperturbed band (see the end of Appendix A in \cite{1}); it means
that the role of evanescent channels is not reflected adequatly. Such
channels  always exist for $d>2$ and give essential contribution
for sufficiently small $L_z$.

 \vspace{5mm}

Few minor remarks.
\vspace{2mm}

(F) It is repeatedly stated in \cite{1} that one-parameter scaling
is proved numerically for many quantities. However, one should be
quite careful with empirical proofs of scaling: it is possible to
suggest an algorithm, which makes it possible to "prove" scaling
in practically any situation \cite{20}[Sec.\,4]. The appearance
of scaling curves presented in Figs.\,53,\,57 in \cite{1}
resembles very closely
the expected result of applying such "algorithm".

\vspace{2mm}

(G) Existence of deviations from scaling is admitted by numerical
researches (see Fig.\,54 in \cite{1}), but it is believed that they
can be correctly analysed. As an example of such analysis, the paper
[133] is cited in \cite{1}, whose results are qualified as
the most accurate and reliable. In fact, the analysis of [133]
arouse serious objections (see discussion in \cite{26}). In our
opinion, the correct interpretation of deviations from scaling is still
an open question.

\vspace{2mm}

{\qquad\qquad\qquad\qquad\qquad--------------------------}

\vspace{2mm}

Overlooking the arguments (A--G), one can see the general tendency:
the numerical algorithms have repeated problems with one-parameter
scaling, on which they are founded. Does it mean a complete failure
of the one-parameter scaling hypothesis? Such possibility can be discussed,
but we prefer a less radical point of view: in principle, scaling exists
but the numerical algorithms do not control a correct choice of scaling
variables.

It looks as a joke, but since 1981 practically nobody studied the
Anderson transition numerically as change in the character of wave
functions. Contemporary numerical algorithms are based on the idea
that any dimensionless quantity $A$ related to a system, spatially
restricted on a scale $L$, is a function of a ratio $L/\xi$ ($\xi$
is the correlation length)
$$
A=F\left(L/\xi \right)\,,
\eqno(2)
$$
which makes it possible to investigate the dependence of $\xi$
on parameters. Relation (2) is a consequence of scale invariance
and is valid under condition that a quantity $A$ has no essential
dependence on microscopical length scales. The latter condition
is difficult to control, since the quantities $A$ used in
numerical algorithms have rather indirect relation to the Anderson
transition.

Let us discuss this point on the example of the ferromagnetic phase
transition. It is well known, that scale invariance is the property
of the spatial picture of fluctuations of the magnetic moment. As
a result, the relation (2) is valid for the quantities that are
directly determined by these fluctuations, i.e. for more or less all
magnetic quantities. If we consider the properties, characterizing
a ferromagnet as a normal metal and related with its electron or
phonon spectra etc., then there is no ground for validity of (2).
Difference between magnetic and nonmagnetic properties is
intuitively evident, but the analogous subdivision becomes nontrivial
for phase transitions of the different nature. One can suggest,
that the numerical algorithms for the Anderson transition use
"nonmagmetic" or not completely "magnetic" quantities. Constructive
analysis of a situation with the Lyapunov exponents is given in
\cite{20}[Sec.\,5].
\vspace{2mm}

{\qquad\qquad\qquad\qquad\qquad--------------------------}

\vspace{2mm}

The author of \cite{1} sees the origin of controversy in the fact
that analytical theory deals with averaged quantities and does not
treat adequately their statistical fluctuations. Such position is
rather weak, as one can see looking at the arguments (A--G).
Indeed,  general scaling philosophy (argument A) does not specify
the exact sense of the Thouless conductance $g$: it is some
characteristic value, which should not be primitively identified with
$\langle g \rangle$ \cite{12}. Of course,  most of analytical
theories deal with average conductivity or density of states, but
self-averaging of these quantities is rigorously proven \cite{27}.
On the other hand, one can consider the higher moments of these
quantities and arrive once again to some variant (more complicated)
of the $\varphi^4$ field theory with the same renormalizability
properties (argument B). The quantity $\beta_{min}$
(arguments C and D) is indeed related
with averaged quantities (the second moments of the Cauchy solution) but
its relation with the self-averaging quantity $\gamma_{min}$ was
discussed in details \cite{20}. The argument E deals directly with
the distribution of $g$. The arguments F and G are related with
the quantities used in numerical algorithms.

Argumentation of \cite{1}, that there is no need to deal with
large systems in order to obtain reliable results, is also weak.
The statement on the stability of the results with the time (during
which the system size $L$ increases essentially) is based on
the rather special choice of publications. In fact, the value of $\nu$
for $d=3$ shows the essential systematic drift with the time:
$\nu=0.66$ \cite{28}, $\nu=1.2\pm 0.3$ \cite{29},
$\nu=1.35\pm 0.15$ \cite{30}, $ \nu =1.45\pm 0.08$ \cite{2},
$ \nu =1.54\pm 0.08$ \cite{31}, $\nu=1.57\pm 0.02$ \cite{32}.
On the other hand, Fig.\,37 in \cite{1} shows the results for 2D systems of
rather large size (till $L=1000$), demonstrating the behavior, which
"naively ... might be interpreted as a metal--insulator transition"
at $W_c\approx 2$. The author of \cite{1} does not accept such "naive"
interpretation and believes that there is no metallic phase in the 2D
case ($W_c=0$). The question arises: why interpretation of the results
for $d=3$ with $L\le 22$ (Fig.\,53),  $d=4$ with $L\le 10$ (Fig.\,61, left),
$d=5$ with $L\le 8$ (Fig.\,61, right) should not be considered as naive?

We should admit, however, that there is one serious argument in favour
of numerical algorithms: investigation of different quantities and
different models leads to approximately the same results for the
critical behavior \cite{1}. When problems with scaling are essential,
such universality looks incredible (though attempts of its explanation
can be made \cite{18}[Sec.\,4.2]). Under the close inspection, this
universality is not so pronounced: one can see from the figure in
\cite{33} that overall scattering of results for the exponent $\nu$
in  the 3D case, related with different methods and models, is
rather large ($1.20\div 1.75$), while the essentially higher
accuracy is declared in the individual experiments
(e.g. $\nu=1.57\pm 0.02$  in \cite{32}).
Recently, the essential counter-example to this universality was discovered:
the use of the quantities $1/\beta_{min}L$ and $1/\gamma_{min}L$ in the
capacity of $A$ gives essentially different results \cite{18}, though
$\beta_{min}$ and $\gamma_{min}$ are very close from the physical viewpoint.
Probably, another examples of such kind can be found, and it may occur that the
discussed universality has a pure psychological origin: the algorithms leading
to the results, essentially different from conventional, are refused
at the early stage of investigation. Historically, such situation
took place with the minimal metallic conductivity, when a lot of
experimentalists during twenty years measured one and the same
"correct" value of it.
\vspace{2mm}

{\qquad\qquad\qquad\qquad\qquad--------------------------}

\vspace{2mm}

In conclusion, we have analysed controversy between the numerical and
analytical results for the critical behavior near the Anderson
transition. It looks, that the most probable reasons for such controversy
are the following: (a) not sufficiently good choice of scaling variables;
(b)  principal inapplicability of one-parameter scaling for $d\ge 4$;
(c) existence of a large length scale $L_0$ for $d=3$;
(d) the use of formulas for multi-channel localization beyond the range
of their applicability.
It is desirable to analyse critically,
what conclusions can be made from the raw data without use of one-parameter
scaling.  The search of the quantities, whose scaling can be rigorously proven,
becomes an urgent problem.

This work is partially supported by RFBR  (grant 03-02-17519).


\end{document}